\documentclass{PoS}

\usepackage{amsmath}
\usepackage{enumitem}
\usepackage{microtype}

\title{Impact of CMS 5.02 TeV dijet measurements on gluon PDFs -- a preliminary view}

\ShortTitle{Impact of CMS 5.02 TeV dijet measurements on gluon PDFs -- a preliminary view}

\author{Kari J. Eskola\\
       University of Jyvaskyla, Department of Physics, P.O. Box 35, FI-40014 University of Jyvaskyla, Finland \\
       Helsinki Institute of Physics, P.O. Box 64, FI-00014 University of Helsinki, Finland \\
       E-mail: \email{kari.eskola@jyu.fi}}

\author{\speaker{Petja Paakkinen}
        \\ University of Jyvaskyla, Department of Physics, P.O. Box 35, FI-40014 University of Jyvaskyla, Finland \\
        E-mail: \email{petja.paakkinen@jyu.fi}}

\author{Hannu Paukkunen\\
       University of Jyvaskyla, Department of Physics, P.O. Box 35, FI-40014 University of Jyvaskyla, Finland \\
       Helsinki Institute of Physics, P.O. Box 64, FI-00014 University of Helsinki, Finland \\
       E-mail: \email{hannu.paukkunen@jyu.fi}}

\abstract{We discuss the implications of the preliminary CMS dijet data from 5.02 TeV pp and pPb collisions for gluon PDFs of the proton and nuclei. The preliminary pp data show a discrepancy with NLO predictions using for example the CT14 PDFs. We find that this difference cannot be accommodated within the associated scale uncertainties and debate the possible changes needed in the gluon PDF. A similar discrepancy is found between the CMS pPb data and NLO predictions e.g. with the EPPS16 nuclear modifications imposed on the CT14 proton PDFs. When a nuclear modification ratio of the pp and pPb data is constructed, the uncertainties in the scale choices and in proton PDFs effectively cancel and a good agreement between the data and EPPS16 is found, except in some bins at backward rapidities corresponding to large $x$ of the nucleus. To assess the impact of these data on the EPPS16 nuclear PDFs, we use a non-quadratic extension of the Hessian PDF reweighting method. A significant reduction in EPPS16 uncertainties is obtained with the fit supporting strong nuclear shadowing and valence-like antishadowing for gluons. We also indicate the possible extensions needed in the EPPS16 parametrization at large $x$.}

\FullConference{XXVI International Workshop on Deep-Inelastic Scattering and Related Subjects (DIS2018)\\
		16-20 April 2018\\
		Kobe, Japan}

\begin{document}

\vspace{-0.1cm}
\section{Introduction and methodology}

Jet production at hadron colliders has proven to be an important process in constraining gluon parton distribution functions (PDFs) both in free-proton fits~\cite{Gao:2017yyd} and more recently also in nuclear PDFs~\cite{Eskola:2016oht}. Reporting on their preliminary data on dijet measurements at 5.02 TeV proton--proton (pp) and proton--lead (pPb) collisions, the CMS collaboration has noticed a discrepancy between their preliminary data and NLO perturbative QCD predictions with various PDFs~\cite{CMS:2016kjd}. Here, using a non-quadratic extension of the Hessian PDF reweighting method~\cite{Paukkunen:2014zia}, we study the impact of these preliminary data first on the CT14~\cite{Dulat:2015mca} proton PDFs and then on the EPPS16~\cite{Eskola:2016oht} nuclear PDFs. We stress that the following considerations are based on preliminary data and need to be refined with the final data, which have been recently presented in Ref.~\cite{Sirunyan:2018qel}.

The Hessian PDF reweighting~\cite{Paukkunen:2014zia} is a method to study the impact of a new set of data on the PDFs, circumventing the need for a fully fledged global analysis. By using suitable approximations, one aims to minimize the figure-of-merit function
\begin{equation}
  \chi^2_\text{new}(\mathbf{z}) = \chi^2_\text{old}(\mathbf{z}) + \sum_{ij}\,(y_i(\mathbf{z}) - y_i^\text{data})\,C^{-1}_{ij}\,(y_j(\mathbf{z}) - y_j^\text{data}),
  \vspace{-0.2cm}
\end{equation}
where $\chi^2_\text{old}$ incorporates our knowledge on the original global analysis in terms of $z_k$, the parameter variations in the eigendirections of the original Hessian matrix around its minimum $\chi^2_0$, $y_i$ describe the theory predictions corresponding to the new experimental input $y_i^\text{data}$, and $C^{-1}_{ij}$ are the elements of the inverse covariance matrix of the new dataset.

\begin{floatingfigure}[tb]
  \includegraphics[width=\textwidth]{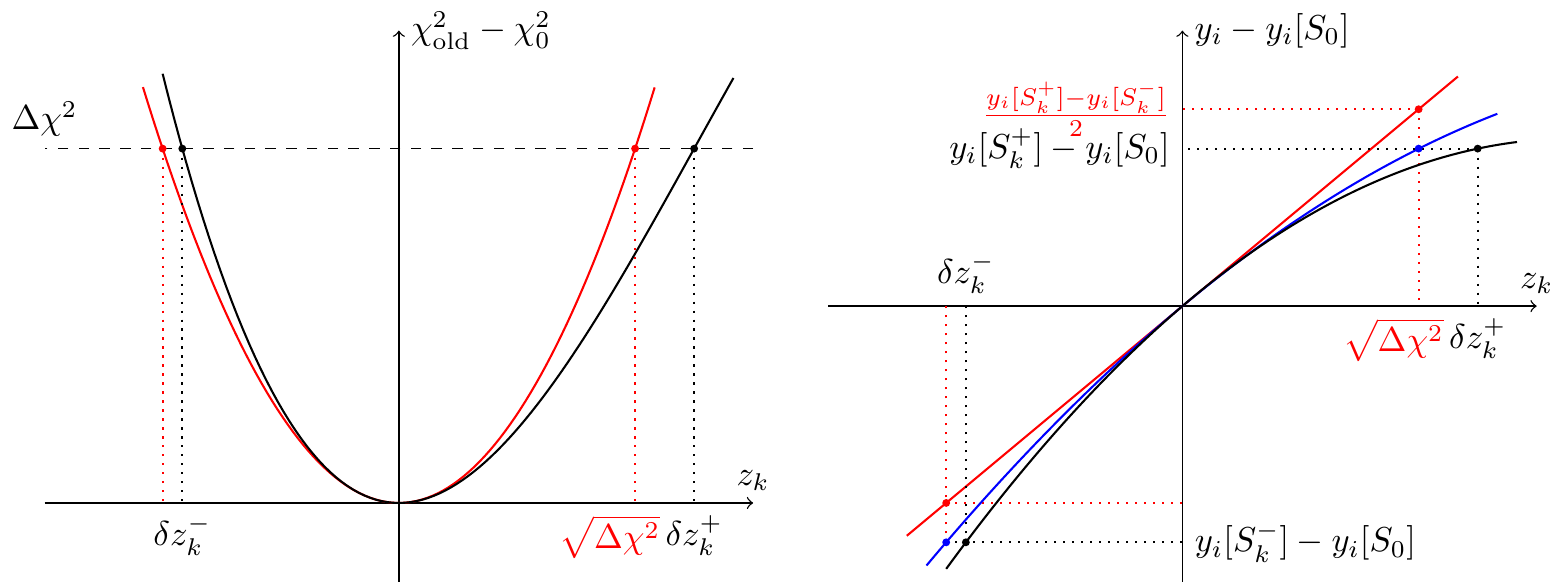}
  \vspace{-0.3cm}
  \caption{Possible approximations for Hessian PDF reweighting: quadratic--linear in red, quadratic--quadratic in blue and cubic--quadratic in black.}
  \label{fig:rw}
\end{floatingfigure}

We illustrate different possible approximations in Figure~\ref{fig:rw}. In the simplest approximation $\chi^2_\text{old}$ is fully quadratic and $y_i$ are just linear functions of $z_k$,
\begin{description}[labelwidth=4cm]
  \item[quadratic--linear:] \makebox[5.5cm]{$\chi^2_\text{old}(\mathbf{z}) \approx \chi^2_0 + \sum_{k} z_k^2$,\hfill} \makebox[6.0cm]{$y_i(\mathbf{z}) \approx y_i[S_0] + \sum_k d_{ik} z_k$,\hfill}
\end{description}
but here we improve the method by taking into account also higher order terms. In particular, we consider the following approximations
\begin{description}[labelwidth=4cm]
  \item[quadratic--quadratic:] \makebox[5.5cm]{$\chi^2_\text{old}(\mathbf{z}) \approx \chi^2_0 + \sum_{k} z_k^2$,\hfill} \makebox[6.0cm]{$y_i(\mathbf{z}) \approx y_i[S_0] + \sum_k (d_{ik} z_k + e_{ik} z_k^2)$,\hfill}
  \item[cubic--quadratic:] \makebox[5.5cm]{$\chi^2_\text{old}(\mathbf{z}) \approx \chi^2_0 + \sum_{k} (a_k z_k^2 + b_k z_k^3)$,\hfill} \makebox[6.0cm]{$y_i(\mathbf{z}) \approx y_i[S_0] + \sum_k (d_{ik} z_k + e_{ik} z_k^2)$.\hfill}
\end{description}
For the quadratic--linear and quadratic--quadratic approximations it suffices to have the PDF error sets and to know the tolerance criterion $\Delta\chi^2$, using which the central prediction $y_i[S_0]$ and the coefficients $d_{ik}$ and $e_{ik}$ can be calculated, but for the cubic--quadratic approximation one needs additional knowledge on how the original $\chi^2_\text{old}$ function deviates from a quadratic one to derive the coefficients $a_{k}$ and $b_{k}$. For EPPS16 this information is provided in Table 2 of Ref.~\cite{Eskola:2016oht}, where $\delta z^\pm_k$, the parameter values at which $\chi^2_\text{old}$ has grown from its minimum by an amount $\Delta\chi^2$, are given.

\begin{floatingfigure}[t]
      \hspace{-0.25cm}\includegraphics[width=0.36\textwidth]{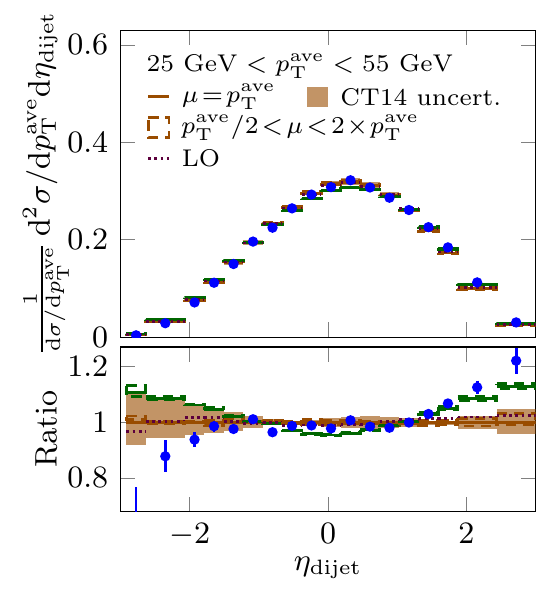}
      \hspace{-0.15cm}\includegraphics[width=0.36\textwidth]{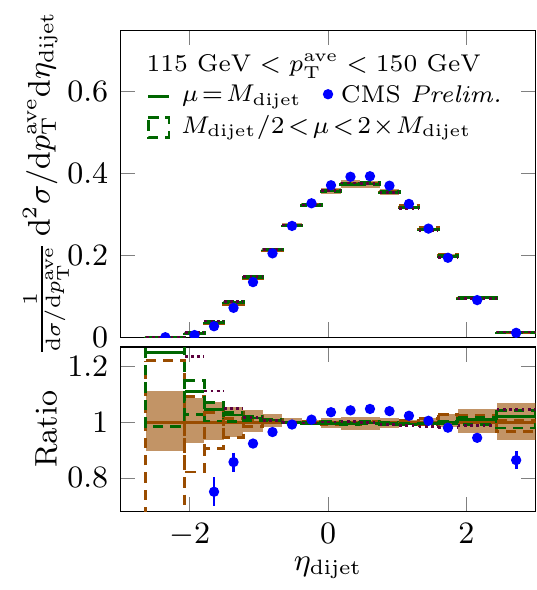}
  \vspace{-0.2cm}
  \caption{The preliminary CMS pp dijet rapidity spectra read off from Ref.~\cite{CMS:2016kjd} compared with NLO predictions using the CT14 PDFs.}
  \label{fig:pp}
\end{floatingfigure}

\begin{floatingfigure}
  \begin{minipage}{0.36\textwidth}
  \hspace{+0.04cm}\includegraphics[width=\textwidth]{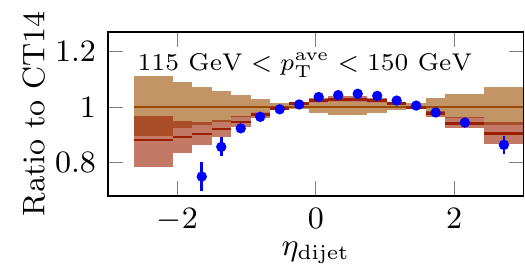}\\
  \hspace{-0.45cm}\includegraphics[width=1.051\textwidth]{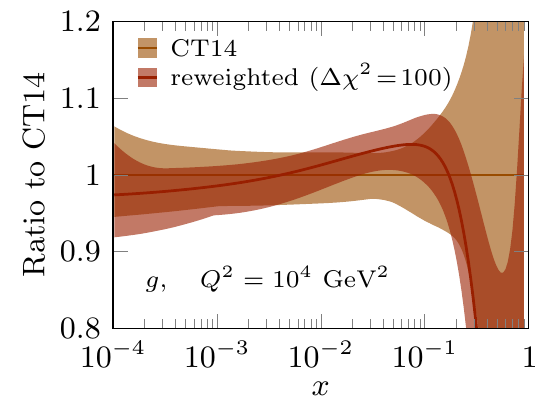}
  \end{minipage}
  \caption{The effect of reweighting on CT14 for the dijet spectrum (top) and gluon PDF (bottom).}
  \label{fig:CT14}
\end{floatingfigure}

\vspace{-0.1cm}
\section{Implications for CT14 proton PDFs}

The preliminary pp rapidity spectra for two bins of dijet average transverse momentum are shown in Figure~\ref{fig:pp}. We compare these data with NLO theory predictions from NLOJet++~\cite{Nagy:2003tz} using the CT14 NLO PDFs~\cite{Dulat:2015mca}. Two scale choices and their variations with factor of two are shown: the scalar average of the transverse momenta of the two jets, $\mu = p_{\rm T}^{\rm ave}$, commonly used for dijets, and the invariant mass of the dijet, $\mu = M_{\rm dijet}$, which has been argued to have a better perturbative convergence~\cite{Currie:2017eqf}. For $\mu = p_{\rm T}^{\rm ave}$ we show also the leading order prediction. We verify the observation in Ref.~\cite{CMS:2016kjd} that the predictions with CT14 are significantly wider than the preliminary data in bins of high $p_{\rm T}^{\rm ave}$. While at the lowest $p_{\rm T}^{\rm ave}$ bin (25 to 55 GeV) the scale-choice effects are sizable throughout the rapidity range, for larger values of $p_{\rm T}^{\rm ave}$ the midrapidity region appears to be robust against scale variations and LO-to-NLO effects. Hence it is unlikely that the discrepancy could be explained just by missing NNLO terms.

Using $\mu = p_{\rm T}^{\rm ave}$ and excluding the lowest-$p_{\rm T}^{\rm ave}$ bin where the scale uncertainty is large, we have performed a reweighting study on the CT14 PDFs. We use here the quadratic--quadratic approximation with $\Delta\chi^2 = 100$, which approximately corresponds to the tolerance in the CT14 error sets. The results are shown in Figure~\ref{fig:CT14}. We find that the reweighting is able to cure the midrapidity discrepancy. For this, an enhancement in the gluon PDF at $x$ around $0.1$ and a suppression at larger $x$ are needed. At $\eta_{\rm dijet} \lesssim -1$ the data are still not reproduced, which might be due to a high-$x$ parametrization issue or NNLO effects as scale uncertainties are large in this region. As it is hard to extract reliably the data uncertainties from Ref.~\cite{CMS:2016kjd}, these results should be considered merely indicative and need to be refined with the final data. We note that the found gluon modifications somewhat resemble those seen when including high-luminosity 7 TeV jet data in the MMHT analysis~\cite{Harland-Lang:2017ytb}.

\begin{floatingfigure}[t]
  \hspace{-0.25cm}\includegraphics[width=0.36\textwidth]{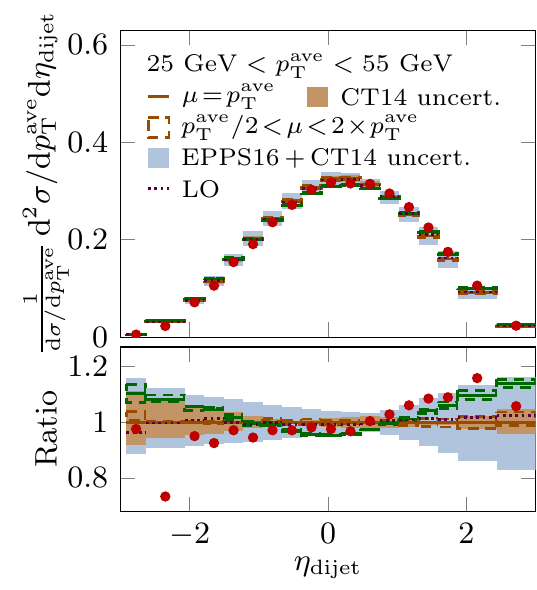}
  \hspace{-0.15cm}\includegraphics[width=0.36\textwidth]{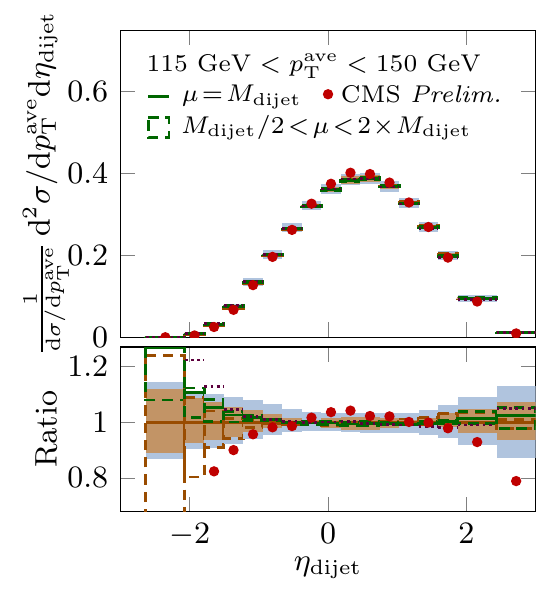}
  \vspace{-0.2cm}
  \caption{The preliminary CMS pPb dijet rapidity spectra read off from Ref.~\cite{CMS:2016kjd} compared with NLO predictions using the CT14 and EPPS16 PDFs.}
  \label{fig:pPb}
\end{floatingfigure}

\begin{floatingfigure}
  \begin{minipage}{0.36\textwidth}
  \hspace{-0.00cm}\includegraphics[width=\textwidth]{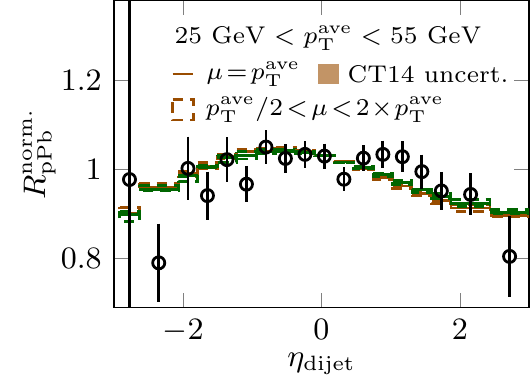}\\
  \hspace{-0.00cm}\includegraphics[width=\textwidth]{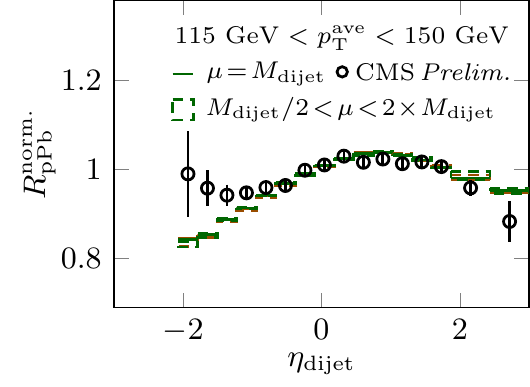}
  \end{minipage}
  \caption{The preliminary CMS data based on Ref.~\cite{CMS:2016kjd} and cancellation of proton-PDF and scale uncertainties in nuclear modification factor.}
  \label{fig:unsc}
\end{floatingfigure}

\vspace{-0.1cm}
\section{Implications for EPPS16 nuclear PDFs}

In Figure~\ref{fig:pPb} we show a comparison of the preliminary pPb rapidity spectra with theory predictions using the CT14 proton PDFs and EPPS16 nuclear modifications. While here the data lie mostly within the combined EPPS16+CT14 uncertainty band, we also observe that the data deviate from the central predictions \emph{the same way} as the pp data do. Thus, as we saw that the preliminary pp data could not be described with the CT14 PDFs without modifications, using the preliminary pPb dijet spectra in a nPDF analysis with CT14 proton PDFs would lead to an overestimation of nuclear effects. For this reason, it is better to use the nuclear modification factor of the normalized differential cross sections
\begin{equation}
  R_{\rm pPb}^{\rm norm.} = \frac{\frac{1}{\mathrm{d}\sigma^{\rm  pPb}/\mathrm{d}p_\mathrm{T}^{\rm ave}}\,\mathrm{d}^2\sigma^{\rm  pPb}/\mathrm{d}p_\mathrm{T}^{\rm ave}\mathrm{d}\eta_{\rm dijet}}{\frac{1}{\mathrm{d}\sigma^{\rm pp}/\mathrm{d}p_\mathrm{T}^{\rm ave}}\,\mathrm{d}^2\sigma^{\rm pp}/\mathrm{d}p_\mathrm{T}^{\rm ave}\mathrm{d}\eta_{\rm dijet}}.
\end{equation}
As seen in Figure~\ref{fig:unsc}, the proton-PDF and scale uncertainties effectively cancel in this observable.

\begin{floatingfigure}[t]
  \hspace{-0.15cm}\includegraphics[width=0.33\textwidth]{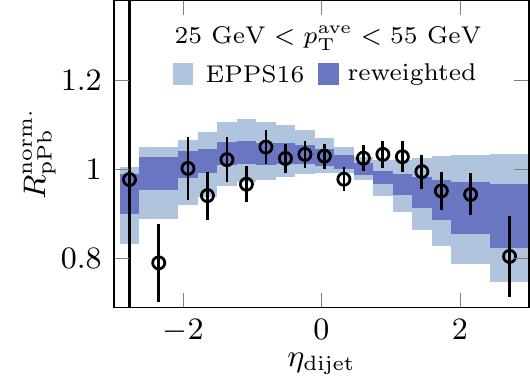}
  \hspace{-0.1cm}\includegraphics[width=0.33\textwidth]{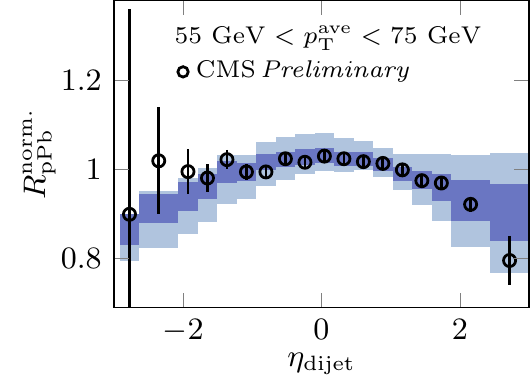}
  \hspace{-0.1cm}\includegraphics[width=0.33\textwidth]{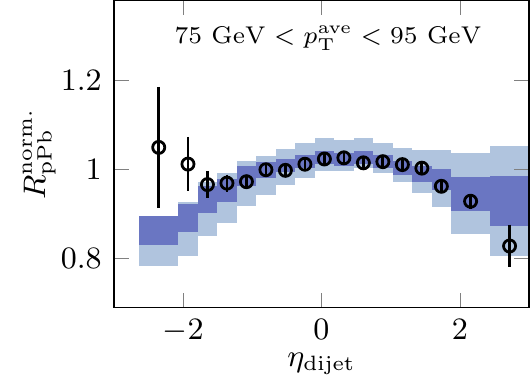} \\
  \hspace{-0.15cm}\includegraphics[width=0.33\textwidth]{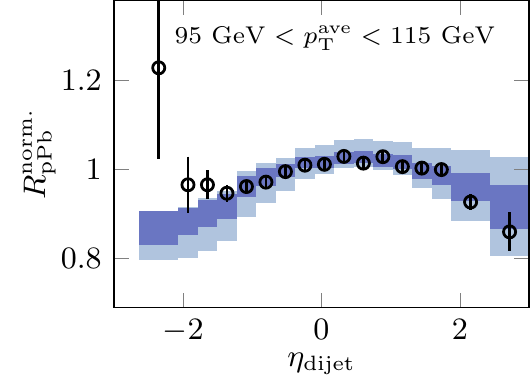}
  \hspace{-0.1cm}\includegraphics[width=0.33\textwidth]{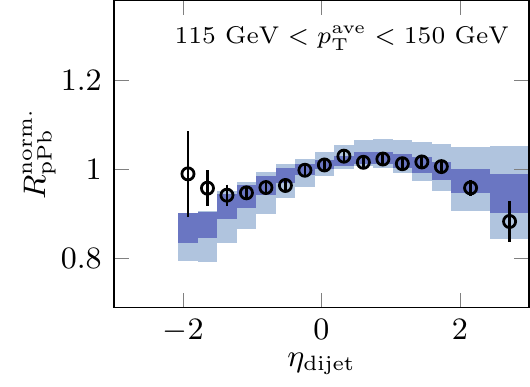}
  \hspace{-0.1cm}\includegraphics[width=0.33\textwidth]{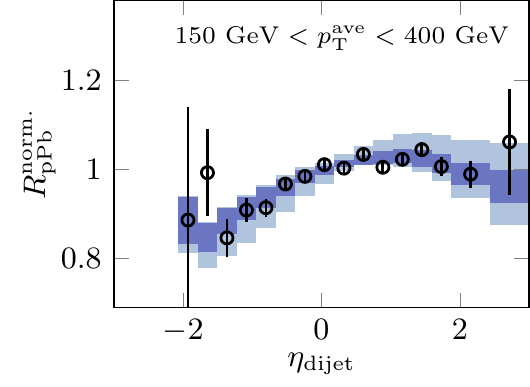}
  \vspace{-0.4cm}
  \caption{The CMS preliminary data for the nuclear modification factor of normalized cross sections read off from Ref.~\cite{CMS:2016kjd}, compared with the EPPS16 predictions and reweighted results.}
  \label{fig:RpPb}
\end{floatingfigure}

We compare the preliminary data for $R_{\rm pPb}^{\rm norm.}$ with the uncertainties from EPPS16 in Figure~\ref{fig:RpPb}. We find the data to be well in line with the EPPS16 predictions except at some bins in backward rapidities corresponding to very high values of $x_{\rm Pb}$. We also see that the data uncertainties are clearly smaller than those of EPPS16, promising a good constraining power. Indeed, performing a reweighting in the cubic--quadratic approximation, the EPPS16 uncertainties shrink to match those of the data.

\begin{floatingfigure}
  \begin{minipage}{0.38\textwidth}
    \includegraphics[width=\textwidth]{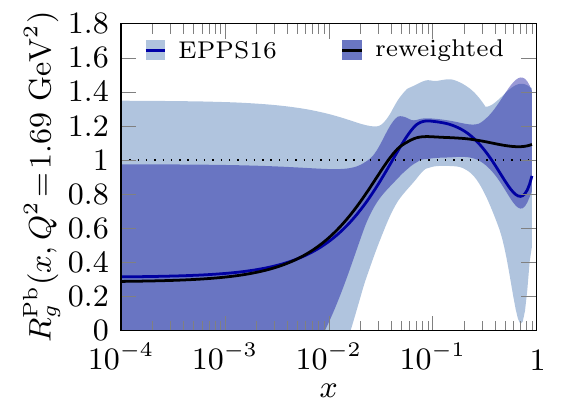}
  \end{minipage}
  \caption{The effect of reweighting with $R_{\rm pPb}^{\rm norm.}$ on EPPS16 nuclear modification of gluon PDF for the lead nucleus.}
  \label{fig:EPPS16}
\end{floatingfigure}

The original and reweighted EPPS16 gluon PDFs of the lead nucleus are shown in Figure~\ref{fig:EPPS16}. We find a drastic reduction in the EPPS16 uncertainties due to inclusion of the new data, especially in the antishadowing region, where the uncertainty is more than halved. Moreover, we find support for small-$x$ gluon shadowing and mid-$x$ antishadowing as the new uncertainty band lies below and above unity, respectively, in these regions. This is in accordance with the findings of Ref.~\cite{Kusina:2017gkz}, where heavy-flavour production at the LHC was used to constrain nuclear gluons. The large-$x$ region appears problematic for reweighting. The preliminary data seem to prefer an EMC pit at smaller $x$ than allowed in EPPS16, where the location is tied to that of valence quarks. Hence the apparent flatness of the reweighted gluon modification might be due to too restrictive a parametrization. However, we note that in the final data~\cite{Sirunyan:2018qel} such a strong upward pull at backward rapidities seems not to be present, and the data could therefore be more easily accommodated within the EPPS16 uncertainties.

\vspace{-0.1cm}
\section{Summary and outlook}

We have considered here the implications that the preliminary CMS 5.02 TeV dijet data~\cite{CMS:2016kjd} have on the gluon component of the CT14 and EPPS16 PDFs, based on Hessian reweighting method. We have observed that the discrepancy between the preliminary pp data and NLO predictions cannot be accommodated within the associated scale uncertainties. We find that possibly large modifications in the CT14 gluon PDF are needed to describe the preliminary data. Particularly, an enhancement in the gluon PDF at $x$ around 0.1 and a suppression at large $x$ are needed.

The dijet pseudorapidity spectra in pPb collisions suffer from a similar discrepancy, which may be due to a need for modifying the proton PDFs. However, we have shown that in the nuclear modification factor of the normalized spectra, the proton-PDF and scale uncertainties effectively cancel and the data are well in line with EPPS16 predictions.

With the precision of the preliminary data, stringent constraints can be put on the gluon modification in lead nucleus, especially in the antishadowing region, where a non-quadratic reweighting procedure yields a drastic reduction in the gluon uncertainty. We find support for small-$x$ gluon shadowing and mid-$x$ antishadowing. The preliminary data seem to prefer a slightly smaller antishadowing than what we have in EPPS16. We find it difficult to describe the preliminary data in some bins at backward rapidities with the EPPS16 nuclear modifications. This may indicate a need for allowing more freedom in the large-$x$ parametrization in future fits. However, this remark might need to be revised with the final version of the CMS dijet data, which have been recently presented in Ref.~\cite{Sirunyan:2018qel}. A refined analysis using the final data will be presented elsewhere.

\acknowledgments

We have received funding from the Academy of Finland, Project 297058 of K.J.E.\ and 308301 of H.P.;
P.P.\ acknowledges the financial support from the Magnus Ehrnrooth Foundation.
The cross-section calculations were performed within a computing cluster of the Finnish IT Center for Science (CSC) under the Project jyy2580.

\end{document}